\documentclass[11pt]{article}

\usepackage[margin=1in]{geometry}
\usepackage[T1]{fontenc}
\usepackage[utf8]{inputenc}
\usepackage{booktabs}
\usepackage{tabularx}
\usepackage{longtable}
\usepackage{array}
\newcolumntype{Y}{>{\raggedright\arraybackslash}X}
\usepackage{hyperref}
\usepackage{xcolor}
\usepackage{graphicx}
\usepackage{float}
\usepackage{amsmath,amssymb}
\usepackage{enumitem}
\usepackage{microtype}
\usepackage[scaled=0.96]{helvet}

\usepackage{setspace}
\usepackage{titlesec}
\usepackage{titling}
\usepackage{tcolorbox}
\tcbuselibrary{skins,breakable}
\usepackage{fancyhdr}
\usepackage{caption}
\usepackage{tikz}
\usepackage{eso-pic}
\usepackage{array}
\usetikzlibrary{arrows.meta,positioning,fit,shapes.geometric,calc}
\usepackage[numbers,sort&compress]{natbib}

\definecolor{prismblue}{HTML}{1F5AA6}
\definecolor{deepnavy}{HTML}{0B1020}
\definecolor{electricblue}{HTML}{2563EB}
\definecolor{cyanaccent}{HTML}{06B6D4}
\definecolor{prismgray}{HTML}{4B5563}
\definecolor{mutedgray}{HTML}{6B7280}
\definecolor{lightgray}{HTML}{F3F4F6}
\definecolor{panelbg}{HTML}{F8FAFC}

\renewcommand{\arraystretch}{1.16}
\setlist{itemsep=0.28em,topsep=0.35em}

\hypersetup{
  colorlinks=true,
  linkcolor=electricblue,
  citecolor=electricblue,
  urlcolor=electricblue,
  pdfauthor={Bin Cao},
  pdftitle={XMatcher: An Open-Source Framework for Powder X-Ray Diffraction Phase Identification}
}

\titleformat{name=\section,numberless}
  {\Large\bfseries\color{deepnavy}}
  {}{0pt}{\MakeUppercase}
  [\vspace{-0.45em}{\color{electricblue!70}\rule{0.18\linewidth}{0.8pt}}]
\titlespacing*{\section}{0pt}{1.8em}{0.85em}
\titleformat{name=\subsection,numberless}
  {\large\bfseries\color{electricblue}}
  {}{0pt}{}
\titlespacing*{\subsection}{0pt}{1.25em}{0.45em}

\tcbset{
  sharp corners,
  boxrule=0.45pt,
  left=1.0em,
  right=1.0em,
  top=0.8em,
  bottom=0.8em,
  colback=panelbg,
  colframe=electricblue!32
}

\renewenvironment{abstract}
  {\begin{tcolorbox}[enhanced,colback=panelbg,colframe=electricblue!40,borderline west={1.6pt}{0pt}{electricblue},title=Abstract,fonttitle=\bfseries\color{deepnavy},coltitle=deepnavy]\small}
  {\end{tcolorbox}}

\renewcommand{\headrulewidth}{0.25pt}
\renewcommand{\headrule}{\hbox to\headwidth{\color{electricblue!30}\leaders\hrule height \headrulewidth\hfill}}
\title{XMatcher: An Open-Source Framework for X-Ray Diffraction Phase Identification}
\author{Bin Cao}
\date{}

\makeatletter
\renewcommand{\maketitle}{%
  \AddToShipoutPictureBG*{%
    \put(\LenToUnit{\paperwidth-2.05cm},\LenToUnit{\paperheight-1.95cm}){%
      \begin{tikzpicture}
        \clip (0,0) circle (0.42cm);
        \node[opacity=0.36,inner sep=0pt] at (0,0)
          {\includegraphics[width=0.84cm,height=0.84cm]{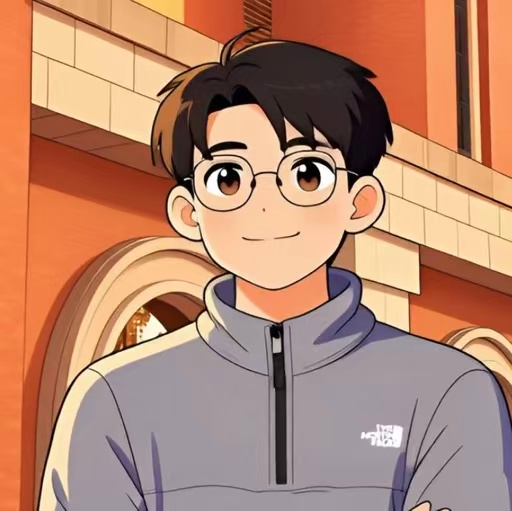}};
        \draw[white,opacity=0.75,line width=0.45pt] (0,0) circle (0.42cm);
      \end{tikzpicture}%
    }%
  }%
  \vspace*{-0.8em}
  \noindent
  \begin{tikzpicture}
    \node[inner sep=0pt,outer sep=0pt,text width=\linewidth] (panel) {\begin{tcolorbox}[
      enhanced,
      colback=white,
      colframe=deepnavy!12,
      boxrule=0.5pt,
      borderline west={2.2pt}{0pt}{electricblue},
      borderline north={0.6pt}{0pt}{cyanaccent!70},
      left=1.25em,right=1.25em,top=1.05em,bottom=1.05em]
      {\footnotesize\bfseries\color{electricblue} XRD PHASE IDENTIFICATION}\par
      \vspace{0.7em}
      {\fontsize{22}{26}\selectfont\bfseries\color{deepnavy}\@title\par}
      \vspace{1.0em}
      {\large\bfseries\color{deepnavy}\@author\par}
      \vspace{0.35em}
      {\small\color{prismgray}The Hong Kong University of Science and Technology (Guangzhou), China \;;\; Huawei Noah's Ark Lab, UK\par}
      {\small\texttt{bcao686@connect.hkust-gz.edu.cn}\par}
      \vspace{0.8em}
      {\scriptsize\color{mutedgray}Phase Retrieval \; \textbullet \; Interpretable Peak Evidence \; \textbullet \; AutoMix \; \textbullet \; Whole Pattern Validation}
    \end{tcolorbox}};
  \end{tikzpicture}
  \vspace{0.8em}
}
\makeatother

\begin{document}

\maketitle

\begin{abstract}
Powder X-ray diffraction (XRD) is widely used for crystalline phase identification, and recent machine learning approaches have demonstrated remarkable capabilities in accelerating diffraction interpretation. However, reliable phase assignment still requires transparent, evidence-based validation, particularly for complex samples where interpretability and expert assessment remain essential. Search-match methods provide a robust and complementary strategy, yet many implementations are proprietary, limiting accessibility and reproducibility.
Here, we introduce XMatcher, an open-source, evidence-driven framework that integrates diffraction databases, matching algorithms, and interactive visualization into a portable workflow. XMatcher generates theoretical diffraction libraries from crystal structures, retrieves candidate phases through chemical and diffraction constraints, applies global angular-shift correction and one-to-one peak matching, and reports quantitative agreement metrics together with peak-level evidence. Its AutoMix module extends identification to multiphase patterns by evaluating candidate phase combinations, estimating non-negative diffraction contributions, and visualizing phase-specific peak distributions.
Through a local graphical interface, XMatcher enables ranked candidate inspection, interactive pattern comparison, PDF/CIF-based whole-pattern validation, and reproducible analysis export. By exposing both supporting and conflicting evidence rather than relying on a single similarity score, XMatcher provides an interpretable and reproducible platform for crystalline phase identification.

\vspace{0.4em}
\noindent\textbf{Keywords:} Powder X-Ray Diffraction; Phase Identification; Search-Match; Multiphase Analysis; Scientific Software.
\end{abstract}
\newpage

\section*{Introduction}
Powder X-ray diffraction remains one of the most widely used techniques for establishing phase identity, monitoring synthesis and diagnosing impurity formation \cite{cao2026physics}. Its practical strength rapid acquisition from polycrystalline material, is also the source of a persistent interpretative difficulty. Distinct structures can share subsets of reflections; measured peak positions can vary because of calibration offsets, specimen displacement, lattice variation or other experimental effects; and multiphase samples can make an apparently convincing single-phase assignment incomplete. Reliable interpretation therefore involves considerably more than selecting the highest similarity score. It requires an unbroken chain of evidence linking the measured profile to candidate structures, theoretical reflections, matched and unmatched peaks, compositional constraints and, for mixtures, phase-specific attribution.

Recent intelligent phase-identification systems \cite{awesome_xrd2crystal_2026} have demonstrated the growing potential of machine learning for pattern recognition in powder diffraction. Current directions include multicategory classification \cite{cao2025xqueryer}, blind source separation \cite{gao2026xdecomposer}, retrieval-based inference \cite{lai2025end} and conditional generation \cite{li2025powder}. Together, these approaches point toward high-throughput and, in some settings, minimally supervised identification workflows \cite{cao2026ai}. They are especially attractive when a large, well-curated training set and a clearly defined label space are available. Nevertheless, their applicability can be limited by domain shift between instruments, sample-preparation conditions, radiation sources and material families; by incomplete or evolving phase libraries; and by the practical need to explain an individual result rather than only predict a category.

For expert analysis, search-and-match retrieval remains a competitive and indispensable approach \cite{warren1990x}. It can incorporate explicit prior knowledge, such as known elements, excluded elements, measurement conditions and expected impurity phases, while allowing the user to inspect the individual reflections that support or contradict a proposed assignment. This expert-in-the-loop character is important for real laboratory samples, where impurity phases, polymorphism, preferred orientation, peak overlap and imperfect databases are routine rather than exceptional. A useful research workflow must therefore answer a direct operational question: \emph{which observations support this candidate, which observations do not, and what should the user inspect next?}

XMatcher was designed around this question. It is a local application and Python toolkit that matches experimental powder-XRD diffraction evidence against theoretical peak libraries derived from crystallographic databases. \textbf{In an era when artificial intelligence is accelerating scientific discovery, the research community increasingly depends on open and shareable scientific infrastructure}: systems in which methodological assumptions, reference data, candidate generation strategies, and supporting evidence can be inspected, reproduced, extended, and improved by others. Although many powerful search-and-match solutions have been developed, a substantial number remain commercial or closed-source, limiting transparent evaluation, local adaptation, methodological innovation, and community-driven development (e.g., JADE \cite{JADE2026}, HighScore \cite{HighScorePlus2026}, Match! \cite{Match2026}, DIFFRAC.EVA \cite{DIFFRACEVA2026}, and PDXL \cite{rigaku2010integrated} ). XMatcher addresses this gap by providing an open platform for search-and-match-based phase identification, enabling reproducible and extensible workflows while preserving the interpretability and expert verification required for reliable crystallographic analysis.

The software combines a fast retrieval stage, a detailed candidate-explanation stage and an optional whole pattern confirmation stage. It filters candidates using available compositional information, estimates a bounded global angular shift, performs one-to-one peak assignment and exposes both supporting matches and contradictory residual evidence. Rather than presenting a candidate as a label alone, XMatcher presents it as an inspectable hypothesis. Version 1.1.0 introduces AutoMix, an automatic multiphase-identification workflow that screens candidate combinations, estimates non-negative component contributions and retains phase-specific theoretical peak locations and relative intensities rather than collapsing a mixture into a single opaque score. The project is openly available at \url{https://github.com/Asterbin/Asterbin-XMatcher}.

\section*{XMatcher workflow}
\begin{figure}[H]
\centering\includegraphics[width=\linewidth]{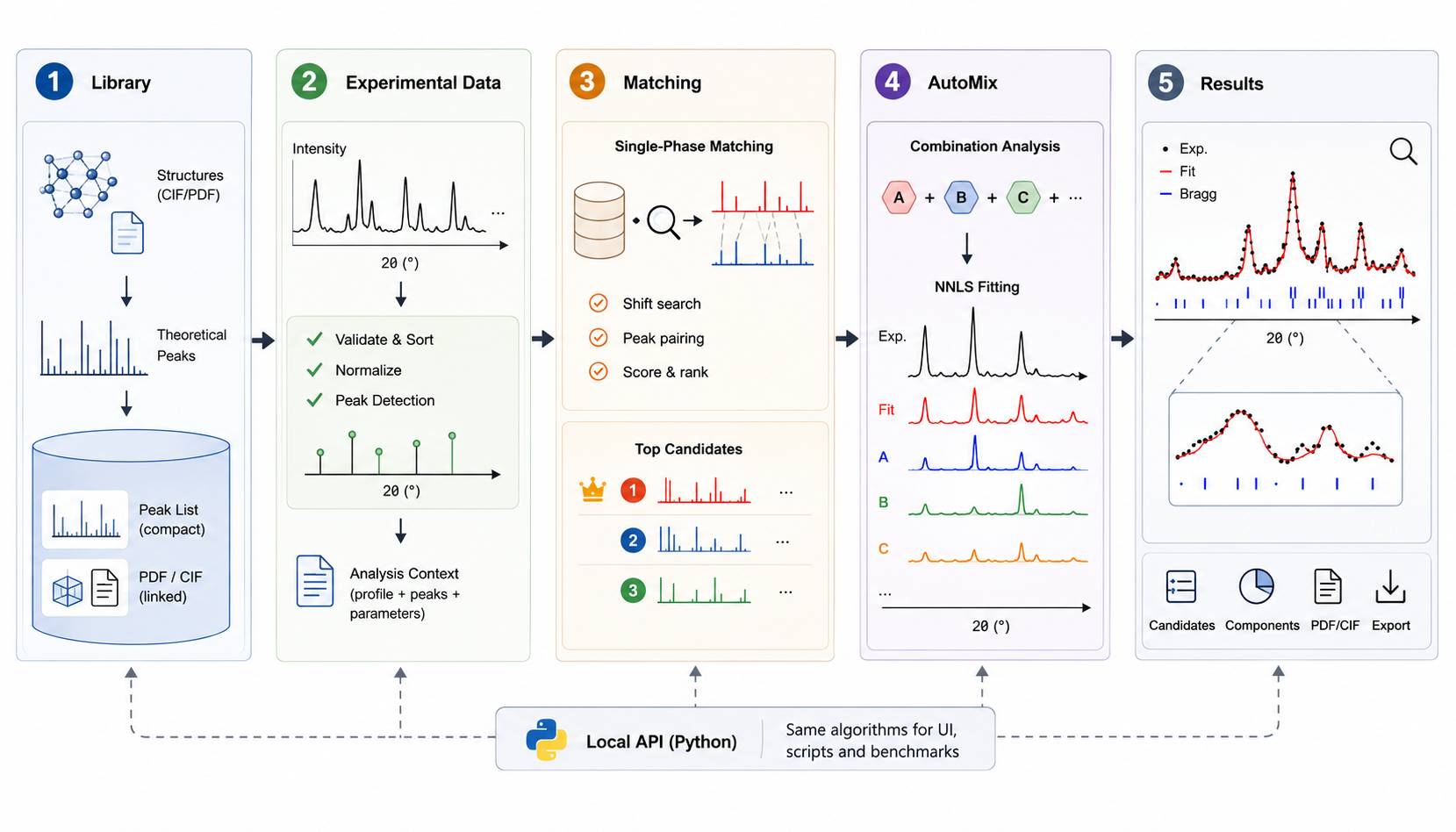}
\caption{\textbf{XMatcher workflow.} Structural data are transformed into a local theoretical peak library. Experimental data are preprocessed and retrieved against the library, then inspected through either single-phase matching, AutoMix combination analysis or whole pattern confirmation.}
\label{fig:workflow}
\end{figure}

XMatcher implements a local, staged processing pipeline from structural records and experimental diffraction data to an inspectable phase-assignment record (Fig.~\ref{fig:workflow}). The workflow starts with an offline library-building process. Structure records, composition metadata and source identifiers are converted into normalized database entries containing elemental sets and theoretical reflection lists. These records are indexed locally, so candidate retrieval does not require transmission of experimental data to a remote service. The resulting library provides two linked representations: a compact peak list for rapid screening and a source-linked PDF/CIF representation for subsequent whole pattern validation.

At analysis stage, the experimental-data module reads either an angle--intensity profile or a user-supplied peak list. It validates numerical columns, orders the angular coordinate, applies the selected intensity normalization and, when a profile is supplied, detects candidate experimental peaks using configurable prominence and separation criteria. The processed peak list, the original profile and the full set of analysis settings are passed to the matching engine as a single analysis context. This prevents the retrieval stage from becoming detached from the measurement and its processing history.

The single-phase engine first applies optional elemental inclusion and exclusion filters to the local library. For each remaining candidate, it searches a bounded global angular offset, evaluates peak-pair eligibility under the specified tolerance and resolves a one-to-one assignment between experimental and theoretical peaks. It then returns a structured result containing the candidate identifier and composition, fitted shift, assigned peak pairs, angular deviations, coverage statistics, missing theoretical peaks and unexplained experimental peaks. The ranking layer consumes these fields to order candidates, while the interface retains the complete fields for inspection rather than receiving only a score.

AutoMix is executed as a second-stage operation on a user-limited pool of leading single-phase candidates. The combination generator enumerates candidate phase sets up to the configured maximum number of components. For each set, theoretical contributions are assembled into a mixture model and non-negative least squares (NNLS) estimates the component weights. Combination-level coverage and residual evidence are calculated alongside the fitted weights. The result object retains the individual identity of every component, its full theoretical peak list and its contribution to the mixture. Because the number of possible combinations grows rapidly with candidate-pool size, the interface exposes this parameter and advises users to keep the pool focused on chemically and diffraction-plausible phases.

The presentation layer consumes the same structured result objects used by the API. For a single phase, it renders candidate summaries, peak-pair evidence and experimental--theoretical overlays. For an AutoMix result, it renders the experimental trace, combined contributions, residual features and separate complete-theory lanes for each component. A user can select alternative combinations, inspect peak metadata and box-select a crowded angular region for magnification. When a candidate requires stronger confirmation, the application requests its linked PDF/CIF information and compares the experimental pattern with the complete theoretical reflection set. 

The implementation separates the browser-facing local interface from the Python matching functions through a local API. The interface therefore orchestrates data selection and visualization without duplicating algorithms, while the same computational functions remain accessible from Jupyter notebooks and scripts. This arrangement supports interactive laboratory use, automated batch analysis and reproducible benchmark execution with the same underlying workflow.

\section*{Features}
\subsection*{Theoretical peak library}
The theoretical library employed in this work is derived from \texttt{MP500.db} \cite{cao2025xqueryer}, comprising 100,315 entries in an ASE-format collection of crystal structures sourced from the Materials Project \cite{jain2013commentary}. The resulting library is distributed as \texttt{MP500\_xrd\_database.pkl} for local analysis. Library construction is performed offline with the \texttt{DatabaseBuilder} pipeline. For each ASE entry, the builder reads the atomic structure and cell, derives the chemical formula and unique element set, retains the Material Project-index identifier, and determines the space-group number and symbol. These fields are stored together with the calculated diffraction representation, so that every retrieval result can be traced back to an MP500 structure record rather than to an unlabelled peak vector.

The diffraction representation is calculated directly from the stored crystal structure. XMatcher converts ASE atoms to a pymatgen structure \cite{ong2013python} and uses \texttt{XRDCalculator} to calculate a powder pattern for the specified radiation. The released build workflow uses Cu K$\alpha$ radiation by default, a $2\theta$ range of $10^{\circ}$--$90^{\circ}$, and excludes reflections below the configured minimum $d$ spacing (default $0.5$ \AA). For each retained reflection, the builder records $2\theta$, normalized relative intensity, reflection family $(hkl)$ and $d$ spacing. Peak intensities are normalized to the strongest calculated reflection of the corresponding structure before the peaks are ranked by intensity.

The \texttt{MP500\_xrd\_database.pkl} release used by XMatcher records a build parameter of thirty strongest calculated reflections per structure, together with wavelength, angular range, minimum $d$ spacing, schema version and entry count in the database metadata. Individual structures can contain fewer retained peaks when fewer valid reflections occur in the configured angular range. A library can therefore be rebuilt with a different peak limit or angular window without changing the matching code. For interactive candidate comparison and AutoMix visualization, XMatcher exposes the complete theoretical peak set available in the selected database entry. Thus, the visible peak lanes are not reconstructed from only the few peaks that happened to contribute to a retrieval score.

The resulting package contains an \texttt{xrd\_database} of peak-bearing MP500 records, an exact element-set index and an inverted element index for contains-mode chemical filtering. The indices are generated from the stored elemental metadata and enable the retriever to reduce the candidate set before numerical peak matching. The package is serialized as a local pickle for rapid loading; because pickle deserialization is unsafe for untrusted files, XMatcher requires that databases be obtained only from trusted project releases or reconstructed from a verified \texttt{MP500.db} source provided through the official repository (\url{https://github.com/Asterbin/Asterbin-XMatcher}).

\subsection*{Single-phase candidate retrieval}
For each theoretical candidate, XMatcher first evaluates chemical admissibility. User-provided required elements define an inclusion condition, and excluded elements remove candidates inconsistent with known sample chemistry. Chemical filtering is optional because unknown contaminants and incomplete composition information are common, but it often transforms a difficult global search into a tractable local comparison.

The experimental and theoretical axes are then aligned with a candidate-level global shift. Rather than independently moving every reflection, XMatcher searches a bounded angular-offset range and selects the shift that maximizes matched evidence under the configured tolerance. This model captures a simple calibration-like offset while avoiding an implausible free fit. The reported shift remains visible to the user; a large shift is therefore an investigation prompt rather than a hidden adjustment.

\begin{figure}[H]
\centering\includegraphics[width=0.85\linewidth]{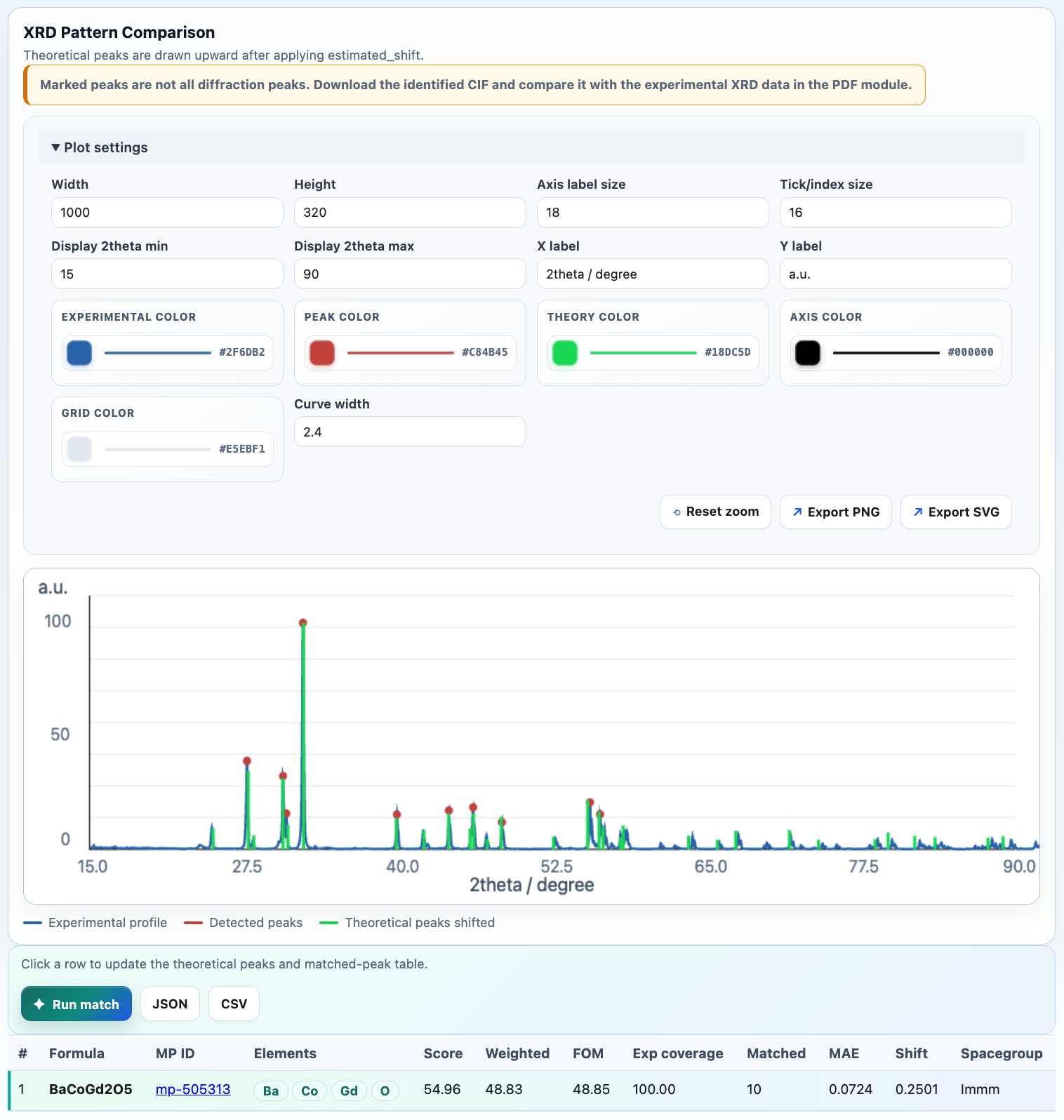}
\caption{\textbf{Single-phase searching.} XMatcher uses a bounded global shift and one-to-one assignment, then reports both the support represented by aligned peaks and the counter-evidence represented by unmatched or missing peaks.}
\label{fig:single}
\end{figure}

Peak assignment is one-to-one (Fig.~\ref{fig:single}). Each experimental peak can support at most one theoretical reflection in a candidate comparison, and each theoretical reflection can be used at most once. This avoids inflating scores when several theoretical lines happen to lie within tolerance of one broad experimental feature. The matching output includes assigned peak pairs, angular deviations, unmatched experimental peaks, and missing theoretical peaks. A composite score is used to rank candidate structures, while its individual components are retained for further analysis.

\subsection*{AutoMix multiphase identification}
AutoMix operates on the same preprocessed experimental profile as single-phase identification. The peak detector first selects the configured number of prominent experimental peaks and reports their $2\theta$ positions and normalized intensities. In unconstrained mode, these peaks are compared with the local theoretical library to form a ranked single-phase candidate pool. The pool is deliberately bounded; the default size is eight candidates and the maximum permitted size is thirty. This limit is computationally important because the number of candidate combinations rises as $\sum_{k=1}^{p}\binom{n}{k}$ for a pool of $n$ candidates and a maximum of $p$ phases. 

Each candidate is represented by a response vector over the detected experimental peaks. A vector element is the theoretical relative intensity assigned to an experimental peak by the one-to-one matcher after its candidate-specific global shift. For a phase combination, these vectors form a matrix $A$, and AutoMix estimates non-negative component coefficients $w$ by solving $\min_{w\geq0}\lVert Aw-y\rVert_2^2$, where $y$ is the experimental peak-intensity vector. The fitted profile is $Aw$; its residual is used to calculate a sum of squared errors, explained intensity and a lightly complexity-penalized combination score. The non-negativity constraint prevents one proposed phase from artificially cancelling another. The returned component values are relative diffraction contributions and are not quantitative mass or volume fractions.

AutoMix also provides a constrained known-phase mode. A user may enter one known chemical component per line as an exact element set, for example \texttt{Na, Cl} and \texttt{Mn, O}. XMatcher resolves each line through the local exact element-set index and retains the best diffraction-supported structure from every matching chemical family. Every accepted combination must contain at least one phase with each requested exact element set. Alternatively, a user may enter one or more MPIDs. Each MPID is resolved to its precise local structure, and every accepted combination must include that structure. Multiple known entries are combined with logical AND: all supplied constraints must be satisfied. This mode avoids an unconstrained scan of irrelevant database structures and is intended for cases where sample chemistry or prior characterization already establishes one or more components. If the supplied constraints cannot be satisfied within the selected maximum phase count, or if no supporting peak evidence is present, AutoMix returns no result rather than relaxing the constraint silently.

\begin{figure}[ht!]
\centering\includegraphics[width=0.9\linewidth]{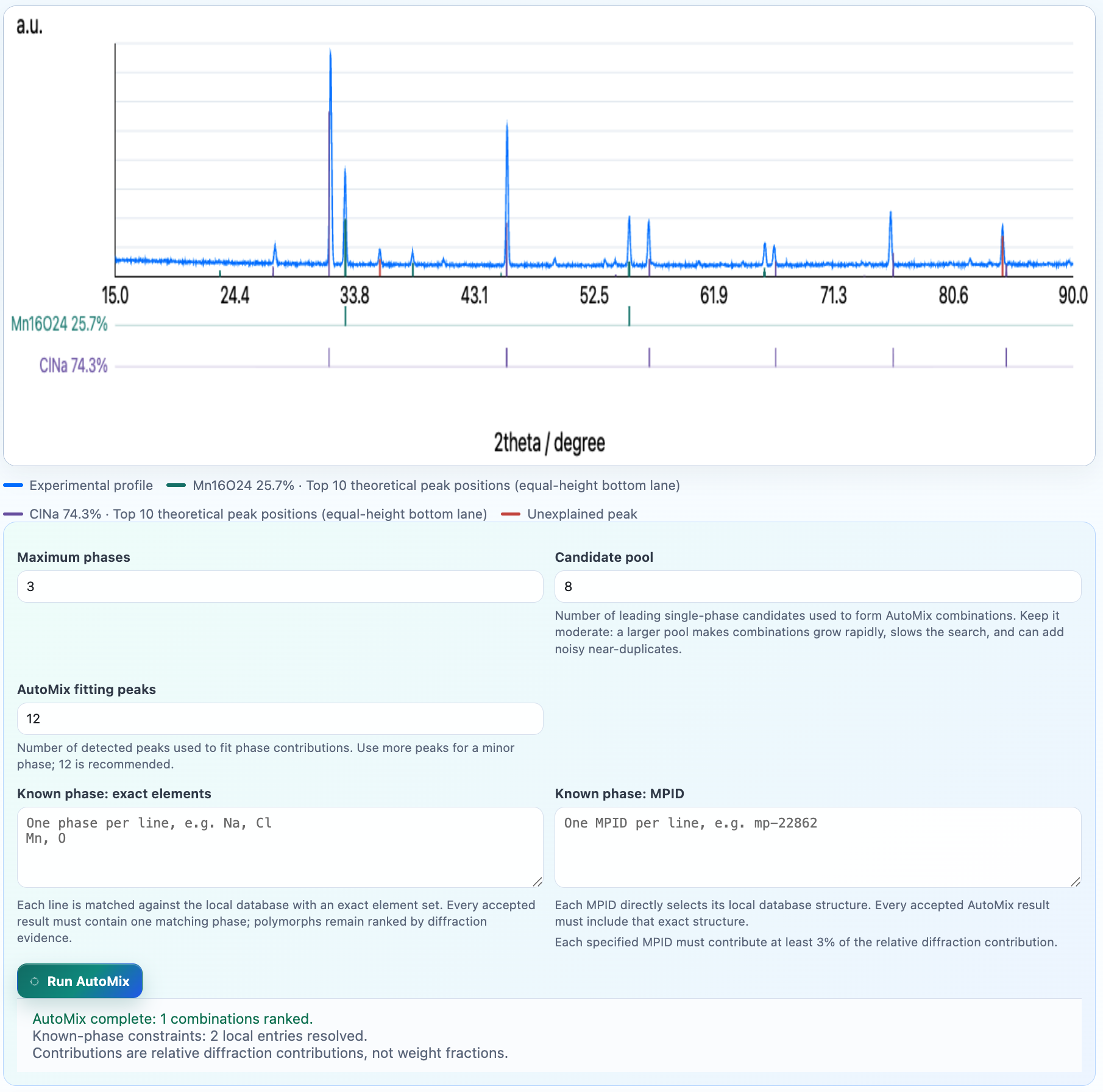}
\caption{\textbf{AutoMix multiphase searching.} AutoMix identifies multiphase compositions by constructing a bounded candidate pool from experimental peak features and theoretical diffraction signatures, then optimizing non-negative phase contributions under unconstrained or user-defined chemical/structural constraints to rank diffraction-consistent phase combinations. }
\label{fig:automix}
\end{figure}

\subsection*{PDF whole pattern comparison}
\begin{figure}[h!]
\centering\includegraphics[width=1\linewidth]{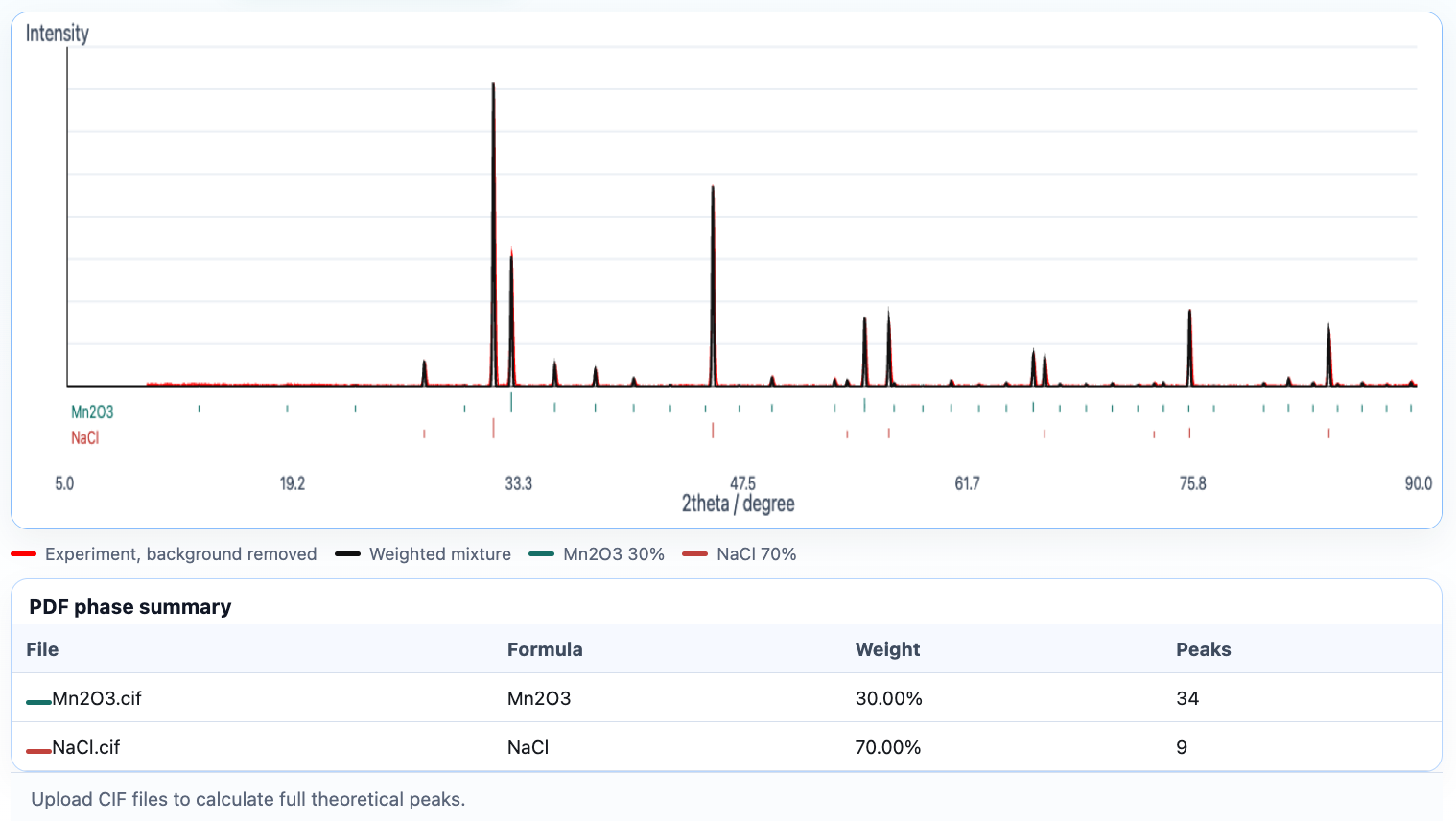}
\caption{\textbf{PDF whole pattern comparison.} The module calculates complete theoretical powder patterns from uploaded CIF structures using the selected radiation, angular range, intensity threshold and peak width. Phase-specific profiles are broadened and combined according to user-defined non-negative relative weights, enabling direct comparison of the simulated mixture with the experimental pattern.}
\label{fig:PDF}
\end{figure}

The PDF (powder diffraction file) whole pattern comparison module is a structure-driven validation workflow that is independent of the compact MP500 retrieval representation (Fig. \ref{fig:PDF}). A user uploads one or more CIF files and provide a dedicated experimental XRD profile. 

For every CIF phase, the local API calculates a theoretical powder pattern with the pysimxrd calculator \cite{cao2025simxrd} the user-selected radiation and $2\theta$ window. The current interface submits Cu K$\alpha$ radiation by default and exposes the lower and upper angular limits, a minimum relative-intensity threshold and a peak-width parameter. Reflections below the selected intensity threshold are omitted from the visual comparison; every retained reflection carries its calculated position, normalized relative intensity, reflection family $(hkl)$ and $d$ spacing. In this sense, the module displays the full calculated reflection set within the explicit angular and intensity settings, rather than only the small subset of peaks used during rapid retrieval.

To support direct mixture inspection, the user assigns a non-negative relative weight to each uploaded phase. XMatcher evaluates every phase on a common $2\theta$ grid, broadens each theoretical reflection with a Gaussian profile whose standard deviation is derived from the selected full width at half maximum, sums the weighted phase profiles and normalizes the resulting mixture profile. The weights are a user-controlled visualization and hypothesis-testing parameter; they are not reported as quantitative phase fractions. A normalization action is provided only to rescale the entered relative weights to a common percentage basis.

\subsection*{Reproducibility and validation}

Reproducible phase identification requires preserving both the computational workflow and the experimental context underlying the analysis. A complete record should include the input diffraction pattern or stable data accession, wavelength and calibration information, database version, preprocessing and peak-detection parameters, composition constraints, matching tolerances, shift ranges, AutoMix settings, and the resulting structured evidence. XMatcher provides validation tests for retrieval and matching procedures, together with checks on programmatic consistency. The principal outputs and their interpretation are summarized in Table~\ref{tab:outputs}.

\begin{table}[H]
\caption{\textbf{Core XMatcher outputs and their interpretation.}}
\label{tab:outputs}
\centering
\footnotesize
\renewcommand{\arraystretch}{1.15}
\begin{tabularx}{\linewidth}{
>{\bfseries\centering\arraybackslash}m{0.20\linewidth}
>{\centering\arraybackslash}X}
\toprule
Output & Interpretation and recommended use\\
\midrule
Candidate rank & Retrieval priority only; inspect score components and peak evidence before assignment.\\
Global shift & Candidate-specific angular offset; investigate large values using calibration and sample context.\\
Matched peak pairs & Direct support for a candidate under the current tolerance and assignment rule.\\
Missing peaks & Counter-evidence; examine especially intense expected reflections.\\
Unexplained peaks & Evidence for impurities, mixture components, artefacts or inadequate library coverage.\\
AutoMix phase lanes & Complete theoretical locations and relative intensities for each selected component.\\
NNLS weights & Non-negative evidence contributions; not quantitative phase fractions without validation.\\
Exported record & Inputs, parameters and evidence required to reproduce an interpretation.\\
\bottomrule
\end{tabularx}
\end{table}

\section*{Discussion}
\subsection*{Open infrastructure}
XMatcher is intended as open scientific infrastructure for a task that is too often mediated by opaque databases, proprietary search engines and irreproducible manual judgement. The central contribution is therefore not a claim that a single score can solve phase identification. It is an openly inspectable architecture in which database records, theoretical reflections, candidate filters, global shifts, one-to-one assignments, residual peaks and multiphase contributions remain accessible to the user. The complete workflow can be run locally, allowing laboratories to analyse proprietary or unpublished measurements without transferring raw patterns to a remote service while retaining control over their structural library and analysis record.

Openness matters at several technical levels. The source code exposes the exact transformations from ASE structure records to the MP500-derived peak library, including radiation, angular range, peak limit and indexing metadata. The matching and AutoMix algorithms can be inspected, tested and replaced without changing the interface contract. The local API makes it possible to integrate the same functions into Jupyter notebooks, batch pipelines or laboratory information systems. A researcher can change the peak detector, scoring function, structure source, mixture model or visualization while retaining a traceable baseline.

\subsection*{Limitations}
XMatcher is a phase-identification support system, not a replacement for crystallographic refinement or chemical judgement. A finite theoretical library cannot recover a phase that is absent from the database. Peak positions can be affected by strain, composition, temperature, specimen displacement and instrumental factors that are not fully represented by a single global shift. Relative intensities can differ substantially because of texture, microabsorption and sample preparation. The AutoMix NNLS weights are evidence weights for a peak representation and should not be interpreted as quantitative phase fractions without a validated calibration and an appropriate profile-refinement model.

\section*{Code availability}
The XMatcher source code, documentation, and release metadata are archived in a permanent repository available at \url{https://github.com/Asterbin/Asterbin-XMatcher}. Standalone applications for Windows and macOS are also publicly available through a version-controlled release archive at \url{https://doi.org/10.6084/m9.figshare.32812985}.

\clearpage

\section*{Methods}

\subsection*{Overview and notation}
XMatcher is a local software workflow that converts an experimental powder-XRD profile into a traceable set of phase hypotheses. The implementation has five connected stages: (i) offline construction of a structure-derived theoretical peak library, (ii) profile validation, baseline treatment and peak detection, (iii) single-phase retrieval with a bounded angular correction and one-to-one peak assignment, (iv) AutoMix combination fitting for multiphase hypotheses, and (v) whole pattern validation from user-supplied CIF/PDF structures. The browser interface communicates with a local Python API; the numerical functions are also callable directly from Python. Thus the same algorithms can be inspected interactively, executed in a Jupyter notebook or included in a batch workflow.

An experimental measurement is represented by ordered observations $D=\{(x_i,I_i)\}_{i=1}^{N}$, where $x_i=2\theta_i$ is the scattering angle (degrees), $I_i\geq0$ is the measured intensity and $N$ is the number of profile samples. A detected-peak representation is $E=\{(e_i,a_i)\}_{i=1}^{m}$, where $e_i$ is the refined position of detected peak $i$, $a_i$ is its non-negative intensity and $m$ is the number of retained experimental peaks. For database phase $q$, the theoretical library stores $T_q=\{(t_{qj},b_{qj},h_{qj},d_{qj})\}_{j=1}^{n_q}$: calculated position $t_{qj}$, calculated relative intensity $b_{qj}$, reflection family $h_{qj}$ and interplanar spacing $d_{qj}$; $j$ indexes the $n_q$ retained reflections. Throughout, $i$ indexes experimental profile samples or peaks, $j$ and $k$ index reflections, and $q$ indexes phases. All quantities returned to the user retain their phase and parameter context.

\subsection*{Theoretical pattern calculation}
For radiation of wavelength $\lambda$, a reflection from planes indexed by Miller indices $(hkl)$ and separated by $d_{hkl}$ occurs at Bragg angle $\theta_{hkl}$ according to Bragg's law,
\begin{equation}
2d_{hkl}\sin\theta_{hkl}=\lambda,
\qquad
2\theta_{hkl}=2\arcsin\left(\frac{\lambda}{2d_{hkl}}\right).
\end{equation}
The structure factor, multiplicity, Lorentz--polarization term and crystallographic symmetry collectively determine an ideal relative intensity. XMatcher delegates this crystallographic calculation to the pymatgen powder-XRD calculator after converting an ASE atomistic record to a pymatgen structure. The calculated intensity is normalized within each phase,
\begin{equation}
b_{qj}^{(\%)}=100\frac{b_{qj}}{\max_k b_{qk}},
\end{equation}
where $k$ ranges over the reflections of phase $q$. Thus $b_{qj}^{(\%)}$ is the intensity of reflection $j$ expressed as a percentage of that phase's strongest calculated reflection. This normalization enables consistent ranking and visualization; it does \emph{not} imply that calculated intensities equal measured intensities, because texture, absorption, extinction, crystallite size, fluorescence, background, microstrain and instrument response can alter a real profile.

\subsection*{MP500 indexing}
The released local library is built offline from \texttt{MP500.db}, an ASE-format structure collection, and saved as \texttt{MP500\_xrd\_database.pkl}. For every valid input record, the builder stores the database entry identifier, MPID, formula, sorted unique element set, atom count, space-group number and symbol, and calculated peak arrays. The released \texttt{MP500\_xrd\_database.pkl} records Cu K$\alpha$ radiation, a $2\theta$ window of $10^{\circ}$--$90^{\circ}$, a minimum $d$ spacing of $0.5$ \AA, and a build limit of the 30 strongest calculated reflections per structure. 

Two composition indexes are constructed. An exact index maps the canonical element tuple $\operatorname{sort}(\mathcal{E}_q)$ to entry identifiers. An inverted index maps every element to the entries containing it. For a query element set $Q$, exact filtering returns $\{q:\mathcal{E}_q=Q\}$, while contains filtering returns $\{q:Q\subseteq\mathcal{E}_q\}$. AutoMix uses a different interpretation for an unconstrained mixture: a component may use any subset of the overall permitted elements, because a mixture containing TiO$_2$ and SiO$_2$ is compatible with the joint element scope $\{\mathrm{Ti,Si,O}\}$ even though neither component contains all three elements.

\subsection*{Experimental peak detection}
Input rows are parsed as numeric angle--intensity pairs, invalid values are removed, negative intensities are clipped to zero and the profile is sorted by increasing $2\theta$. The raw profile is preserved separately from the display-processed profile. For visualization and peak detection, XMatcher applies configurable smoothing and an estimated baseline. Let $\tilde I_i$ denote the smoothed intensity at sample $i$ and $B_i$ the local baseline estimate at the same angle; the non-negative processed signal is
\begin{equation}
I_i^{\prime}=\max\left(0,\tilde I_i-B_i\right).
\end{equation}
Local maxima are accepted subject to a minimum height, prominence and angular separation. If adjacent sampling points around a selected maximum are available, the reported maximum may be refined locally. The detector returns peaks in descending intensity order; retrieval uses the configured leading $m$ peaks. The maximum normalization used for matching is
\begin{equation}
a_i^{(\%)}=100\frac{a_i}{\max_{r\in\{1,\ldots,m\}}a_r}.
\end{equation}
Here the denominator is the largest retained experimental-peak intensity, so $a_i^{(\%)}\in[0,100]$. The detector settings are part of the analysis record because a minor phase can disappear from a peak-only search if too few experimental peaks are retained. 

\subsection*{Single-phase searching}
\label{Single-phase searching}
An experimental pattern and a theoretical entry can differ by a small common zero offset. For every phase $q$, XMatcher evaluates a bounded shift $\delta\in[-\Delta,\Delta]$ (degrees), where $\Delta$ is the user-specified maximum absolute shift, using a regular grid plus experimentally/theoretically implied pair differences. A potential experimental--theoretical pair $(i,j)$ is eligible only if
\begin{equation}
|e_i-(t_{qj}+\delta)|\leq\tau,
\end{equation}
where $\tau$ (degrees) is the positional tolerance. Its cost combines normalized angular error and relative-intensity disagreement,
\begin{equation}
C_{ij}(\delta)=w_x\frac{|e_i-(t_{qj}+\delta)|}{\tau}
+w_I\frac{|a_i^{(\%)}-b_{qj}^{(\%)}|}{100},
\qquad w_x+w_I=1.
\end{equation}
where $w_x\geq0$ and $w_I\geq0$ are the positional and intensity weights, respectively. Ineligible pairs have infinite cost. A linear-sum assignment is then solved only on rows and columns containing at least one finite pair. This detail is important: an unexplained experimental peak is residual evidence and must not make the valid assignments for the rest of a candidate infeasible. The selected assignment $M_q(\delta)$ is one-to-one, so no broadened experimental feature can be reused to support multiple theoretical reflections.

For each shift, XMatcher calculates precision $P=|M_q(\delta)|/m$, theoretical recall $R=|M_q(\delta)|/n_q$, matched experimental-intensity coverage $C_E$, and matched theoretical-intensity coverage $C_T$. Thus, $P$ is the fraction of retained experimental peaks that are assigned and $R$ is the fraction of theoretical reflections that are recovered. The latter two are intensity-weighted fractions,
\begin{equation}
C_E=\frac{\sum_{(i,j)\in M_q(\delta)}a_i^{(\%)}}{\sum_i a_i^{(\%)}},
\qquad
C_T=\frac{\sum_{(i,j)\in M_q(\delta)}b_{qj}^{(\%)}}{\sum_j b_{qj}^{(\%)}}.
\end{equation}
The mean pair cost determines a quality factor $Q=100\exp[-\operatorname{mean}_{(i,j)\in M_q(\delta)}C_{ij}]$. The default hybrid ranking score is
\begin{equation}
H=0.45\,Q\left(\frac{P+R}{2}\right)+0.30\,(100C_T)+0.15\,(100C_E)+0.10\,[100\min(P,R)].
\end{equation}
Here $C_E$ and $C_T$ are fractions, whereas $Q$ and $H$ are expressed on a 0--100 scale. The score orders hypotheses; it does not certify phase identity. The full pair list, shift, missing theoretical peaks and unmatched experimental peaks are retained for inspection.

\subsection*{AutoMix searching}
\label{AutoMix searching}
Without prior phase information, AutoMix uses the leading single-phase candidates to construct all combinations from one phase to the selected maximum. For $n$ candidates and maximum size $p$, the count is $\sum_{k=1}^{p}\binom{n}{k}$; the default candidate pool is $n=8$, giving 92 combinations for $p=3$. This controlled pool prevents a combinatorial expansion from obscuring the result and limits the computation to phases with initial diffraction support.

Known-phase mode changes the candidate construction deliberately. One exact element set per line, such as \texttt{Na, Cl}, is resolved through the exact MP500 element index. One MPID per line is resolved directly to a local entry. All supplied constraints use logical AND: every accepted combination must include a phase with every requested exact element set and every explicitly requested MPID structure. The highest-scoring candidate from each required element family is retained even if its score is lower than unconstrained candidates. In known-phase mode, AutoMix restricts its pool to these resolved structures instead of scanning unrelated database entries. Consequently, known chemical information both accelerates the calculation and remains an explicit, auditable condition on the output.

For a candidate combination containing $r$ phases, XMatcher builds a response matrix $A\in\mathbb{R}^{m\times r}$. Its entry $A_{i\ell}$ is the normalized calculated intensity attributed by the shifted, one-to-one match from component phase $\ell$ to detected peak $i$; it is zero when that component has no assigned reflection at that position. Each non-zero column is scaled to a maximum of one. Given the experimental peak vector $y=(a_1,\ldots,a_m)^\mathsf{T}$, coefficients are obtained by non-negative least squares,
\begin{equation}
\hat w=\arg\min_{w\geq0}\left\|Aw-y\right\|_2^2.
\end{equation}
where $w=(w_1,\ldots,w_r)^\mathsf{T}$ is the vector of non-negative fitted scale coefficients and $\|\cdot\|_2$ is the Euclidean norm. The fitted signal and residual vector are $\hat y=A\hat w$ and $\varepsilon=y-\hat y$, respectively. The residual sum of squares is $\mathrm{SSE}=\sum_i \varepsilon_i^2$, and the unpenalized fit score is
\begin{equation}
V=100\max\left(0,1-\frac{\mathrm{SSE}}{\sum_i y_i^2}\right).
\end{equation}
For ranking, AutoMix uses $S=V-0.75(r-1)$, where the second term is a deliberately small penalty for each additional component. For component $\ell$, the displayed relative diffraction contribution is
\begin{equation}
\rho_\ell=100\frac{\sum_i A_{i\ell}\hat w_\ell}{\sum_i\hat y_i}.
\end{equation}
The denominator is the total fitted signal across the retained peaks, so the non-zero $\rho_\ell$ values sum to 100\%. These values describe relative contribution to the selected peak evidence only. They are not Rietveld-refined phase fractions and must not be interpreted as mass or volume percentages.

Peak attribution assigns a detected peak to the phase with the largest fitted component at that position. A peak is marked as overlapping when more than one phase makes a substantial contribution. A residual peak is an observation for which $y_i>\hat y_i$. For an explicitly required MPID, AutoMix rejects a combination unless its relative diffraction contribution exceeds the configured minimum (default 3\%); this prevents a required phase from being retained merely as a numerical zero.

\subsection*{PDF comparison}
The PDF whole pattern comparison module provides a profile-level validation independent of the truncated retrieval representation. One or more CIFs are parsed as non-primitive structures and recalculated for the chosen radiation and angular interval. For each retained reflection $k$, a Gaussian broadening model is evaluated on a common angular grid $x$,
\begin{equation}
g_k(x)=I_k\exp\left[-\frac{1}{2}\left(\frac{x-x_k}{\sigma}\right)^2\right],
\qquad \sigma=\frac{\mathrm{FWHM}}{2\sqrt{2\ln2}}.
\end{equation}
Here $x_k$ and $I_k$ are the calculated position and relative intensity of reflection $k$, $\sigma$ is the Gaussian standard deviation, and FWHM is the user-specified full width at half maximum (all angular quantities are in degrees). With user-specified non-negative display weights $v_q$ for phase $q$, the simulated mixture is $G(x)=\sum_qv_q\sum_{k\in q}g_k(x)$ and is normalized for overlay. A minimum-intensity cutoff, FWHM, angular range and weights are all shown in the interface and included in the API response. This module is a structured plausibility check: missing strong calculated reflections, systematic displacement and persistent measured residuals are counter-evidence, while positional coincidence alone is not a quantitative refinement.

\clearpage

\bibliographystyle{unsrtnat}
\bibliography{references}

@phdthesis{cao2026physics,
  title={Physics-Constrained Learning of Crystal Structures and Properties from Powder Diffraction},
  author={Cao, Bin},
  year={2026},
  school={The Hong Kong University of Science and Technology}
}

@misc{awesome_xrd2crystal_2026,
  author       = {{awesome-xrd2crystal contributors}},
  title        = {Awesome XRD $\rightarrow$ Crystal},
  year         = {2026},
  howpublished = {\url{https://github.com/Bin-Cao/awesome-xrd2crystal}},
  note         = {Data set}
}

@article{cao2025xqueryer,
  title={XQueryer: an intelligent crystal structure identifier for powder X-ray diffraction},
  author={Cao, Bin and Zheng, Zinan and Liu, Yang and Zhang, Longhan and Wong, Lawrence WY and Weng, Lu-Tao and Li, Jia and Li, Haoxiang and Zhang, Tong-Yi},
  journal={National Science Review},
  volume={12},
  number={12},
  pages={nwaf421},
  year={2025},
  publisher={Oxford University Press}
}

@article{gao2026xdecomposer,
  title={XDecomposer: Learning Prior-Free Set Decomposition for Multiphase X-ray Diffraction},
  author={Gao, Hanyu and Cao, Bin and Su, Yunyue and Zhang, Tong-Yi and Liu, Qiang},
  journal={arXiv preprint arXiv:2605.05866},
  year={2026}
}

@article{li2025powder,
  title={Powder diffraction crystal structure determination using generative models},
  author={Li, Qi and Jiao, Rui and Wu, Liming and Zhu, Tiannian and Huang, Wenbing and Jin, Shifeng and Liu, Yang and Weng, Hongming and Chen, Xiaolong},
  journal={Nature Communications},
  volume={16},
  number={1},
  pages={7428},
  year={2025},
  publisher={Nature Publishing Group UK London}
}

@article{lai2025end,
  title={End-to-end crystal structure prediction from powder X-Ray diffraction},
  author={Lai, Qingsi and Xu, Fanjie and Yao, Lin and Gao, Zhifeng and Liu, Siyuan and Wang, Hongshuai and Lu, Shuqi and He, Di and Wang, Liwei and Zhang, Linfeng and others},
  journal={Advanced Science},
  volume={12},
  number={8},
  pages={2410722},
  year={2025},
  publisher={Wiley Online Library}
}

@inproceedings{cao2025simxrd,
  title={SimXRD-4M: big simulated X-ray diffraction data and crystal symmetry classification benchmark},
  author={Cao, Bin and Liu, Yang and Zheng, Zinan and Tan, Ruifeng and Li, Jia and Zhang, Tong-yi},
  booktitle={International Conference on Learning Representations},
  volume={2025},
  pages={70721--70745},
  year={2025}
}

@book{warren1990x,
  title={X-ray Diffraction},
  author={Warren, Bertram Eugene},
  year={1990},
  publisher={Courier Corporation}
}

@misc{Match2026,
  author       = {{Crystal Impact GbR}},
  title        = {{Match!: Phase Identification Using Powder Diffraction Data}},
  year         = {2026},
  howpublished = {\url{https://www.crystalimpact.com/match/}},
  note         = {Accessed: 2026-07-17}
}

@misc{JADE2026,
  author       = {{Materials Data Inc.}},
  title        = {{JADE Pro}},
  year         = {2026},
  publisher    = {Materials Data Inc.},
  address      = {Livermore, CA, USA},
  howpublished = {\url{https://www.icdd.com/mdi-jade/}}
}

@misc{HighScorePlus2026,
  author       = {{Malvern Panalytical}},
  title        = {{HighScore Plus}},
  year         = {2026},
  publisher    = {Malvern Panalytical B.V.},
  address      = {Almelo, The Netherlands},
  howpublished = {\url{https://www.malvernpanalytical.com/en/products/category/software/x-ray-diffraction-software/highscore}}
}

@misc{DIFFRACEVA2026,
  author       = {{Bruker AXS GmbH}},
  title        = {{DIFFRAC.EVA}},
  year         = {2026},
  publisher    = {Bruker AXS GmbH},
  address      = {Karlsruhe, Germany},
  howpublished = {\url{https://www.bruker.com/en/products-and-solutions/diffractometers-and-x-ray-microscopes/x-ray-diffractometers/diffrac-suite-software/diffrac-eva.html}}
}

@article{rigaku2010integrated,
  title={Integrated X-ray powder diffraction software PDXL},
  author={Rigaku, Corporation},
  journal={Rigaku J},
  volume={26},
  pages={23--27},
  year={2010}
}

@article{jain2013commentary,
  title={Commentary: The Materials Project: A materials genome approach to accelerating materials innovation},
  author={Jain, Anubhav and Ong, Shyue Ping and Hautier, Geoffroy and Chen, Wei and Richards, William Davidson and Dacek, Stephen and Cholia, Shreyas and Gunter, Dan and Skinner, David and Ceder, Gerbrand and others},
  journal={APL materials},
  volume={1},
  number={1},
  year={2013},
  publisher={AIP Publishing}
}

@article{ong2013python,
  title={Python Materials Genomics (pymatgen): A robust, open-source python library for materials analysis},
  author={Ong, Shyue Ping and Richards, William Davidson and Jain, Anubhav and Hautier, Geoffroy and Kocher, Michael and Cholia, Shreyas and Gunter, Dan and Chevrier, Vincent L and Persson, Kristin A and Ceder, Gerbrand},
  journal={Computational Materials Science},
  volume={68},
  pages={314--319},
  year={2013},
  publisher={Elsevier}
}

@article{cao2026ai,
  title={AI-Driven Structure Refinement of X-ray Diffraction},
  author={Cao, Bin and Zhang, Qian and Feng, Zhenjie and Zhang, Taolue and Huang, Jiaqiang and Weng, Lu-Tao and Zhang, Tong-Yi},
  journal={arXiv preprint arXiv:2602.16372},
  year={2026}
}

\end{document}